\documentclass{aa}  
\usepackage{booktabs,caption,fixltx2e}
\usepackage[flushleft]{threeparttable}
\usepackage{graphicx}
\usepackage{grffile}
\usepackage{natbib}
\bibpunct{(}{)}{;}{a}{}{,} 

\usepackage{txfonts}
\usepackage{xcolor}
\begin{document}

   \title{A Deconvolution Technique to Correct Deep Images of Galaxies from Instrumental Scattered Light}

   \author{E. Karabal\inst{1,2}
          \and
          P.-A. Duc\inst{1}
          \and
          H. Kuntschner\inst{2}
          \and
          P. Chanial\inst{3}
          \and
          J.-C. Cuillandre\inst{1,4}
          \and
          S. Gwyn\inst{5}
          }

   \institute{Laboratoire AIM Paris-Saclay,  CEA/Irfu/SAp, CNRS, Universit\'e Paris Diderot, 91191 Gif-sur-Yvette Cedex, France
         \and
             European Southern Observatory, Karl-Schwarzschild-Str. 2, 85748 Garching, Germany
         \and
         Astroparticule et Cosmologie (APC), CNRS-UMR 7164, Universit\'e Paris 7 Denis Diderot, 10, rue Alice Domon et L\'eonie Duquet, F-75205 Paris Cedex 13, France 
         \and
             Observatoire de Paris, PSL Research University, France
         \and
         	NRC, National Research Council of Canada, 5071 West Saanich Rd, Victoria, British Columbia, V9E 2E7, Canada
             }

   \date{Accepted December 14, 2016}

   \abstract{Deep imaging of the diffuse light emitted by the stellar
     fine structures and outer halos around galaxies is now often used
     to probe their past mass assembly.  Because the extended halos
     survive longer than the relatively fragile tidal features, they
     trace more ancient mergers. We use images reaching surface
     brightness limits as low as 28.5--29
     mag.arcsec\textsuperscript{-2} (g--band) to obtain light and
     color profiles up to 5--10 effective radii of a sample of nearby
     early-type galaxies. They were acquired with MegaCam as part of
     the CFHT MATLAS large programme. These profiles may be compared
     to those produced by simulations of galaxy formation and
     evolution, once corrected for instrumental effects.  Indeed they
     can be heavily contaminated by the scattered light caused by
     internal reflections within the instrument. In particular, the
     nucleus of galaxies generates artificial flux in the outer halo,
     which has to be precisely subtracted.  We present a deconvolution
     technique to remove the artificial halos that makes use of very
     large kernels. The technique based on PyOperators is more time
     efficient than the model-convolution methods also used for that
     purpose. This is especially the case for galaxies with complex
     structures that are hard to model.  Having a good knowledge of
     the Point Spread Function (PSF), including its outer wings, is
     critical for the method. A database of MegaCam PSF models
     corresponding to different seeing conditions and bands was
     generated directly from the deep images.
     It is shown that the difference in the PSFs in different bands
     causes artificial changes in the color profiles, in particular a
     reddening of the outskirts of galaxies having a bright nucleus.
     The method is validated with a set of simulated images and
     applied to three representative test cases: NGC\,3599, NGC\,3489,
     and NGC\,4274, and exhibiting for two of them a prominent ghost
     halo. The method successfully removes it.
       
 }

 
 \maketitle
%

\section{Introduction} \label{Intro}

The mass growth and morphological evolution of galaxies is described
by different scenarios in the literature: gas accretion from
cosmological filaments, internal secular evolution and series of
galaxy mergers of different mass ratios and gas fractions.  Depending
on the dominant scenario, the simulations modeling these processes
predict specific observables for the resulting galaxy, such as stellar
age and metallicity profiles probed by spectroscopy as well as light
and color profiles traced by photometry
\citep[e.g.][]{naab07,bullock05,oser10,cooper10,pillepich15}.
 
While the inner regions of galaxies have been well studied, the
extended parts are still an immature but rapidly growing field of
research.  Multiple programs are being carried out to explore these
regions that hold the imprints of recent and more ancient mergers.
They make use of either stellar counts for local galaxies, or the
diffuse light revealed by deep imaging for more distant ones
\citep{mihos05,delgado10,janowiecki10,abraham14,duc15,iodice16}. New
imaging techniques have been developed
\citep{mihos05,ferrarese12,vandokkum14}, allowing us to reach the very
low surface brightness levels needed to investigate the extended
stellar structures of galaxies. Deep images reveal the fine structures
produced by recent mergers (shells, streams, tails), but also the
outer stellar halos which cumulate material accreted during more
ancient collisions. Stellar halos in extended regions survive longer
than the fine structures. Unfortunately their photometric analysis
suffers from various sources of contamination.

The difficulties of deep imaging should be taken care of elaborately
to reach the desired sensitivity in the extended parts of light and
color profiles. Various contamination sources hamper the integrated
light analysis of the halos, in particular galactic cirrus and
instrumental contamination.  Due to its complex filamentary structure
\citep{miville16}, and non-monochromatic colours, the optical emission
from Galactic cirrus cannot be easily subtracted; heavily cirrus
polluted regions should thus be avoided. Instrumental contamination
affects all fields; some can be addressed by advanced flat-fielding
procedures and sky subtraction; others like reflections within the
camera, producing artificial halos and ghosts, are much more complex
to model and remove \citep{slater09,duc15}.  The importance of
correcting for light scattering is well documented
\citep{michard02,slater09,sandin15, trujillo15} but the correction
process is challenging. It requires in particular the availability of
point spread functions (PSFs) measured up to large radius at the
location of the target on the detector.  Minimising artificial halos
should in principle be done at the instrumental level. The scattering
in classical cameras may be reduced with anti-reflective coating on
the detector and on the optical elements \citep{mihos05} or largely
eliminated by alternative imaging systems such as photo-lens array
\citep{abraham14}.

\citet{sandin14,sandin15} has investigated the impact of ghost halos
on galaxy profiles by convolving galaxy models with PSFs corresponding
to a variety of instruments. He quantified the importance of
scattering as a function of radius for various morphological types.
\cite{duc15} noted the wavelength dependence of the scattering effect,
which for the MegaCam instrument, climax in the r band. The variation
of the PSF shape from one band to the other creates artificial
reddening in the outer color profiles of galaxies.  \citet{trujillo15}
applied a model convolution technique to estimate and remove light
scattering: a pre-determined intrinsic galaxy model is convolved with
the PSF, and the best solution is iteratively found fitting the
convolved model to the data. With this method, having a
multi-component model that minimises residuals in the model-subtracted
image is fundamental: positive or negative residuals cause
under-estimation or over-estimation of the scattered light. Building
interactively a specific multi-component galaxy model is time
consuming and becomes prohibitive for large surveys of massive,
complex, galaxies.

In this paper, instead of using the model-convolution technique, we
present our efforts to directly deconvolve images and remove
artificial halos of galaxies. The deconvolution method relies on large
kernels and uses PyOperators\footnote{http://pchanial.github.io/pyoperators/} \citep{chanial12}, which are time-efficient python
tools. Our analysis is based on CFHT/MegaCam deep images acquired as
part of the MATLAS (Mass Assembly of early-Type GaLAxies with their
fine Structures) large programme \citep{duc15} as well as simulated
images used to validate the method.
 
This paper is structured as follows: in Section \ref{Method}, we
present the MATLAS images and characterize their ghost halos; we
describe the method to determine the point spread functions to large
radii, detail the deconvolution technique and the routines to derive
light and color profiles. In Section \ref{Results}, we present the
results obtained with images of simulated galaxies and apply the
method for images of three real galaxies -- NGC\,3489, NGC\,3599 and
NGC\,4274 -- that exhibit a variety of artificial halos.  In Section
\ref{Discussion}, we discuss the limitations of the method due to
uncertainties on the PSF and potential issues dealing with saturated
images. In Section \ref{Conclusions}, the conclusions follow.

\section{Data and method} \label{Method}

\subsection{The ghosts on MATLAS images} \label{OriginofGhosts}

MATLAS is a deep imaging survey that uses MegaCam on the CFHT
\citep{boulade03} to acquire images of nearby early-type galaxies from
the sample of the ATLAS$^{\sc 3D}$ project \citep{cappellari11}. The
ATLAS$^{\sc 3D}$ sample is volume limited (within 42~Mpc) and includes
galaxies with dynamical mass in the range
${\sim}$~$10^{\text{9.6}}$-$10^{\text{11.8}}$ $M_{\odot}$
\citep{cappellari2013a}.

MATLAS images are obtained by stacking 6--7 individual images, each of
which has 6 minutes exposure time. The individual pointings are made
with large offsets from one another (in the range of 2-14 arcmin) to
enable the detection of extended low-surface-brightness features.  The
images are collected in the g'- and r'-band and additionally some
galaxies are observed in u* and i'\footnote{Noted hereafter, u, g, r and
  i.}. In most cases, the target galaxies are located close to the
center of the 1x1 square degree stacked image.  Galaxies that are
within the field of an already observed object were not re-observed;
some are thus located closer to the edge at locations where the shape
of the PSF is significantly changed with respect to that measured in
the central regions.  Further details on the observing strategy and
data reduction technique can be found in \citet{duc15}.

\begin{figure}[t]
\includegraphics[width=8.8cm]{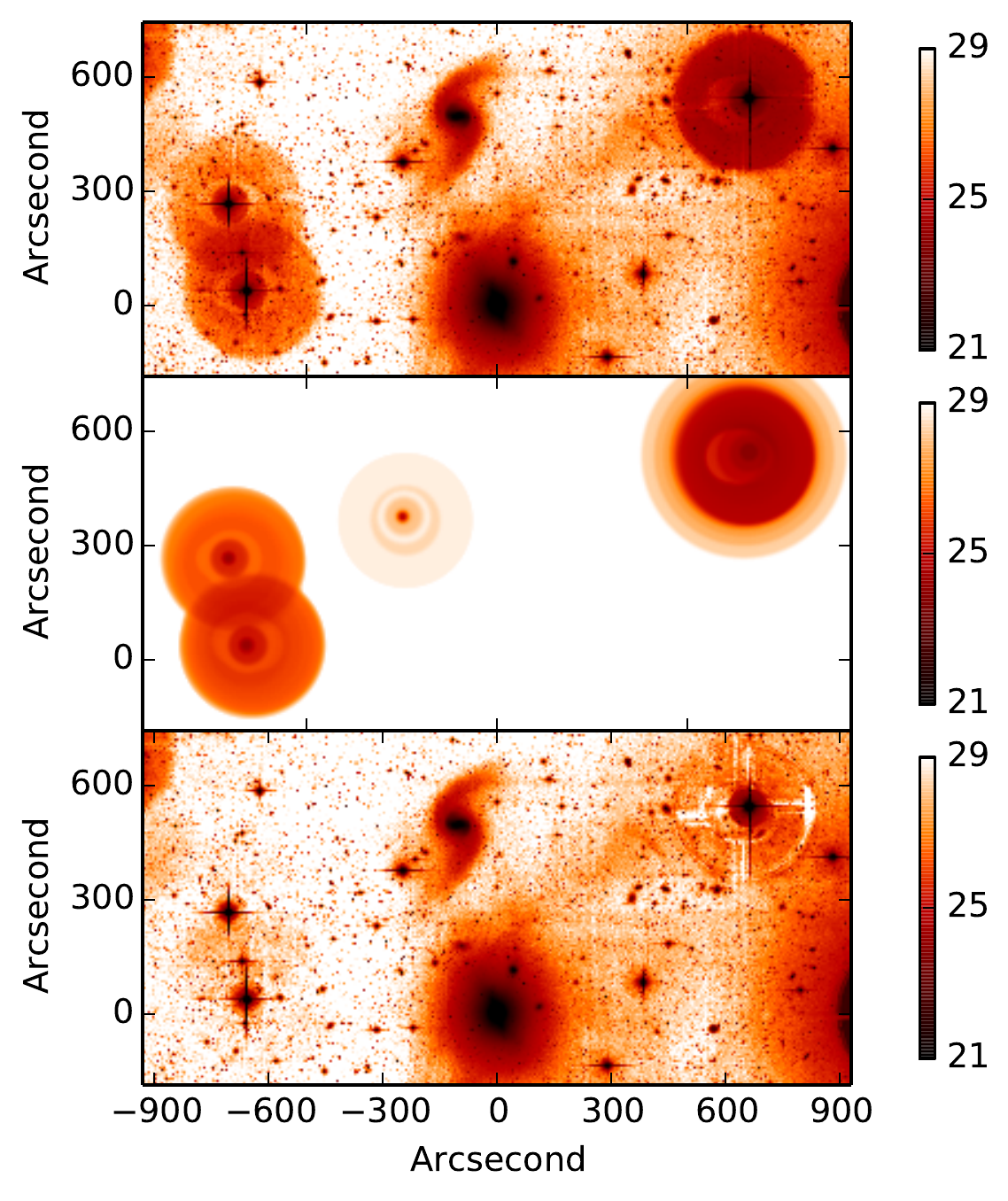}
\centering
\caption{The surface brightness maps in r-band of the stellar ghosts
  around NGC\,3414 \textit{(top)}, the ghost models \textit{(middle)}
  and the ghost subtracted image \textit{(bottom)}. The color bar to
  the right represents the surface brightness level. The target galaxy
  is centred at the [0,0] coordinate of the map. The maps are blurred
  with a Gaussian (sigma = 0.55\arcsec) to make faint regions visible.}
\label{fig:StellarGhost}
\end{figure}

The MATLAS images were usually obtained under good seeing conditions
and benefit from the excellent image quality of the MegaCam camera,
while reaching low local surface brightness limits (28.5-29
mag.arcsec\textsuperscript{-2} in the g-band). They are however
subject to prominent scattering effect due to the complex optical
structure of the instrument. Multiple internal reflections create
characteristic extended interleaved halos (from now referred as
ghosts) around bright objects, stars, but also galaxies.

The ghosts are due to the light coming to the camera, that is being
reflected back and forth between the CCD and the other optical
elements such as the lenses and filter wheels. The origin and the
physical processes behind these ghosts were investigated by
\citet{slater09}. The presence of ghosts alters the photometric
analysis, i.e. change light and color profiles, especially in the
outskirts of galaxies, beyond the $\gtrsim$ 24
mag.arcsec\textsuperscript{-2} isophote. The radius of the most
prominent ghost seen in the MegaCam deep images is $\sim$3.5 arcmin.

As illustrated in Figure~\ref{fig:StellarGhost},  ghosts are present everywhere in the images, contributing to the image
background; they are however directly visible only around bright
objects. The two sources of ghosts -- the bright stars and galactic nuclei
-- cause external and self-contamination, respectively. Different methods are applied
to remove them; the ghosts of bright stars can be easily identified on the images, manually modeled and subtracted.  
 On the other hand,  reflection halos surrounding galaxies are more complex as the light contributed to them is extended. Besides  such ghosts 
 cannot  be easily identified and separated from the stellar halo of the  galaxy, especially when the latter is extended. Therefore, only a
  deconvolution technique can properly remove them.
 Note that this process does not eliminate the ghosts around the bright stars since they are saturated.

\subsection{Subtraction of the ghosts around the bright stars close to the galaxies} \label{SubofStel}
The extended halos around the bright stars, close to the target galaxies,  contaminate their stellar halos, and need to be subtracted.

\begin{figure}[t]
\includegraphics[width=8.8cm]{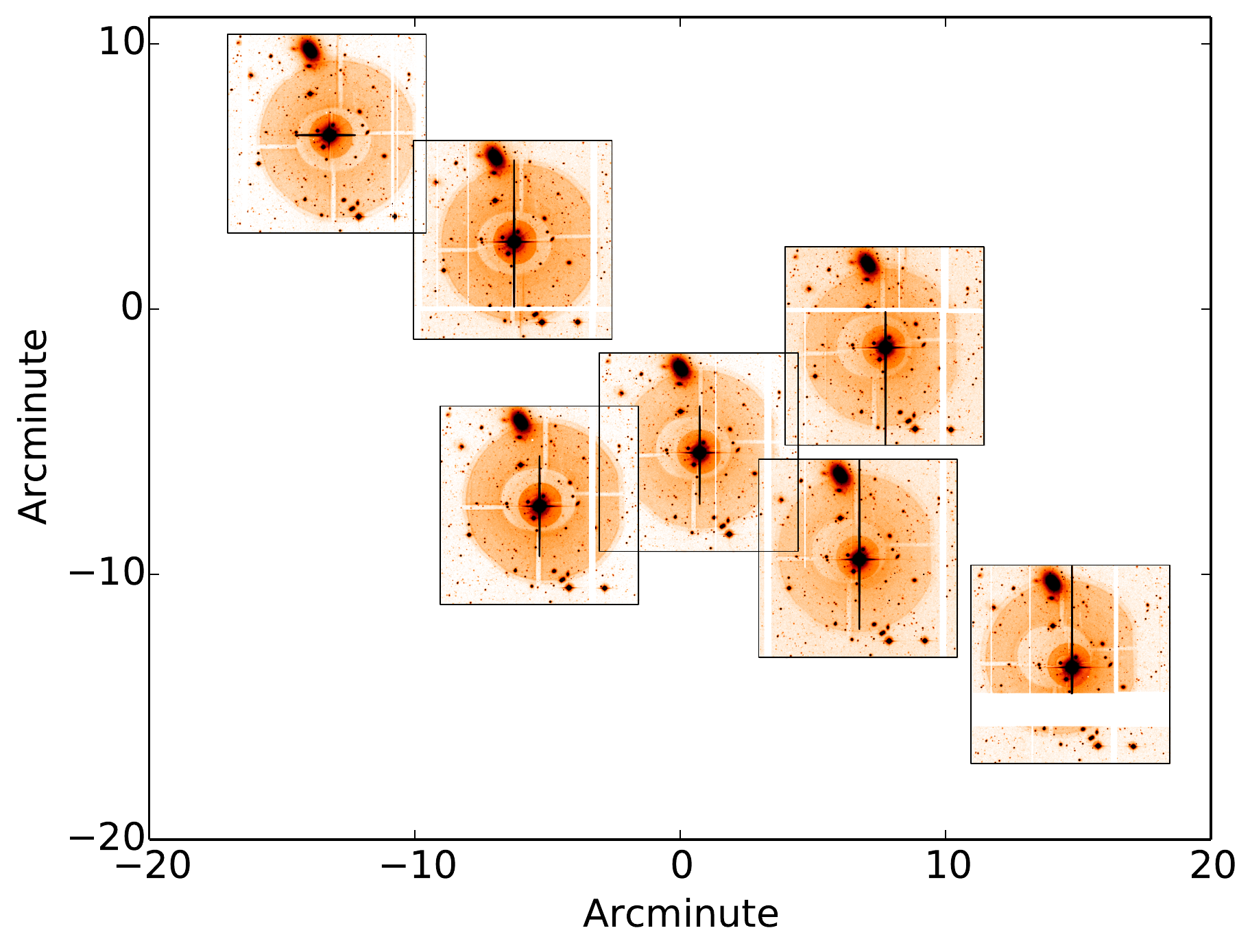}
\centering
\caption{Example of ghosts around a bright star close to the galaxy PGC\,028887 (located to the North of the star). Each stamp corresponds to one of the 7  individual  images obtained for that field and  centered on the star. The positions of the  stamps on the main panel  reflect the observation strategy used for the MATLAS survey, making large offsets between the individual images. Note how the position of the main reflection annulus with
respect to the star changes from one image to the other.  }
\label{fig:GhostPositions}
\end{figure}

\begin{figure}[t]
\includegraphics[width=8.8cm]{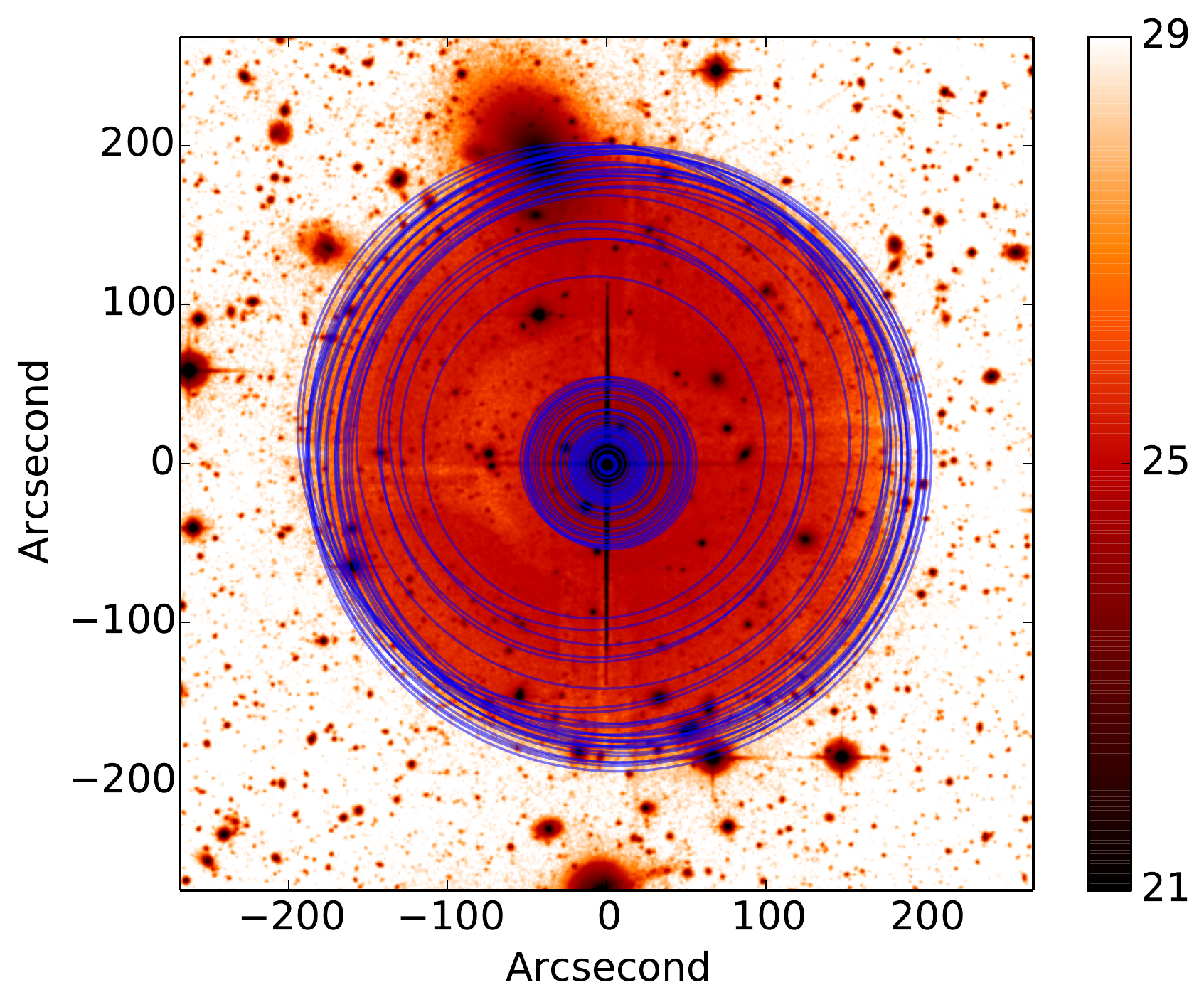}
\centering
\caption{The ghost halo of the star shown in Figure~\ref{fig:GhostPositions} as seen in the stacked image. The blue circles correspond to the boundary of the individual disks used for the manual modeling of the halo. }
\label{fig:Reflections}
\end{figure}

On individual images, these ghosts  consist of multiple annuli partially overlapping each other that, depending on their position on the CCDs, and bands used,  have  different angular sizes, centres and brightnesses.
The stacking procedure  of  these individual images that have large mutual spatial offsets  (see Figure~\ref{fig:GhostPositions})  adds to their complexity. 
A manual interactive procedure was then set up to generate models of the stacked halos: series of annuli with manually adjusted values were
constructed, starting with the external regions. Each annulus is
subtracted from the image and the residual image is iteratively
processed until the halo disappears and a flat background is
reached. The procedure is  illustrated in   Figure~\ref{fig:Reflections}. Note that the method  is not able to properly model the gaps with lower intensity that are present between some annuli especially in the individual images. On the stacked images, they are however partly filled (see Figure~\ref{fig:Reflections}) and  modeling them  would anyway be too complex. 
The light fitting  procedure starts at a radius of 200\arcsec, and is stopped at $\sim$50\arcsec. Modeling the profiles further in would not help much for the purpose of decontaminating the galaxy  halo by the stellar ghost.

This process creates sharp edges while the boundary of the original
ghosts were smoother, due to the stacking procedure on images obtained
with large offsets.  To take this into account, a gaussian blur was
applied on the final ghost model.  The shape of the halos
is wavelength dependent, as further discussed in
Section~\ref{PSFProfiles}.  Thus different models had to be
constructed for each band.  

Finally, the ghost halo models were subtracted from the images, as illustrated in  Figure~\ref{fig:StellarGhost}  for the field of galaxy NGC\,3414.

\subsection{Determination of the Point Spread Function} \label{ObtainPSF}
 \label{PSFProfiles}

 Deconvolution techniques such as the one presented here require in
 principle a perfect knowledge of the Point Spread Function at the
 location, time and wavelength of the observations, and to determine
 it from the innermost to the outermost regions.  In the optical
 regime, and without any AO facility, the shape of the inner PSF is at
 first order governed by the seeing of the atmosphere. Far out, its
 outer wings are shaped by the instrumental, and especially for
 MegaCam by the multiple, thousands, internal reflections.

 Building such a PSF is not straightforward. A physical modeling of
 the PSF, taking into account the complex light path between the
 different optical elements of the camera, can be envisioned by ray
 tracing techniques for instance. However, it would be computationally
 expensive to determine such PSF models for different locations on the
 image and for different wavelength ranges. Instead, empirical methods
 are being used, and the PSF is directly derived from the science
 exposures. One way is to take numerous short exposure images,
 preventing saturation, and stack them until the signal-to-noise in
 the outer part of the PSF is high enough to model the wings
 \citep{janowiecki10,trujillo15}. Such a calibration requires
 dedicated observations, matching the seeing conditions of the science
 observations, and thus a lot of telescope time.  Alternatively, one
 may stack numerous non-saturated stars from one or if needed several
 science exposures, but this process would smooth the non-symmetric
 features of the PSF. Indeed they vary with position and time.

\begin{figure}[t]
\includegraphics[width=9.8cm]{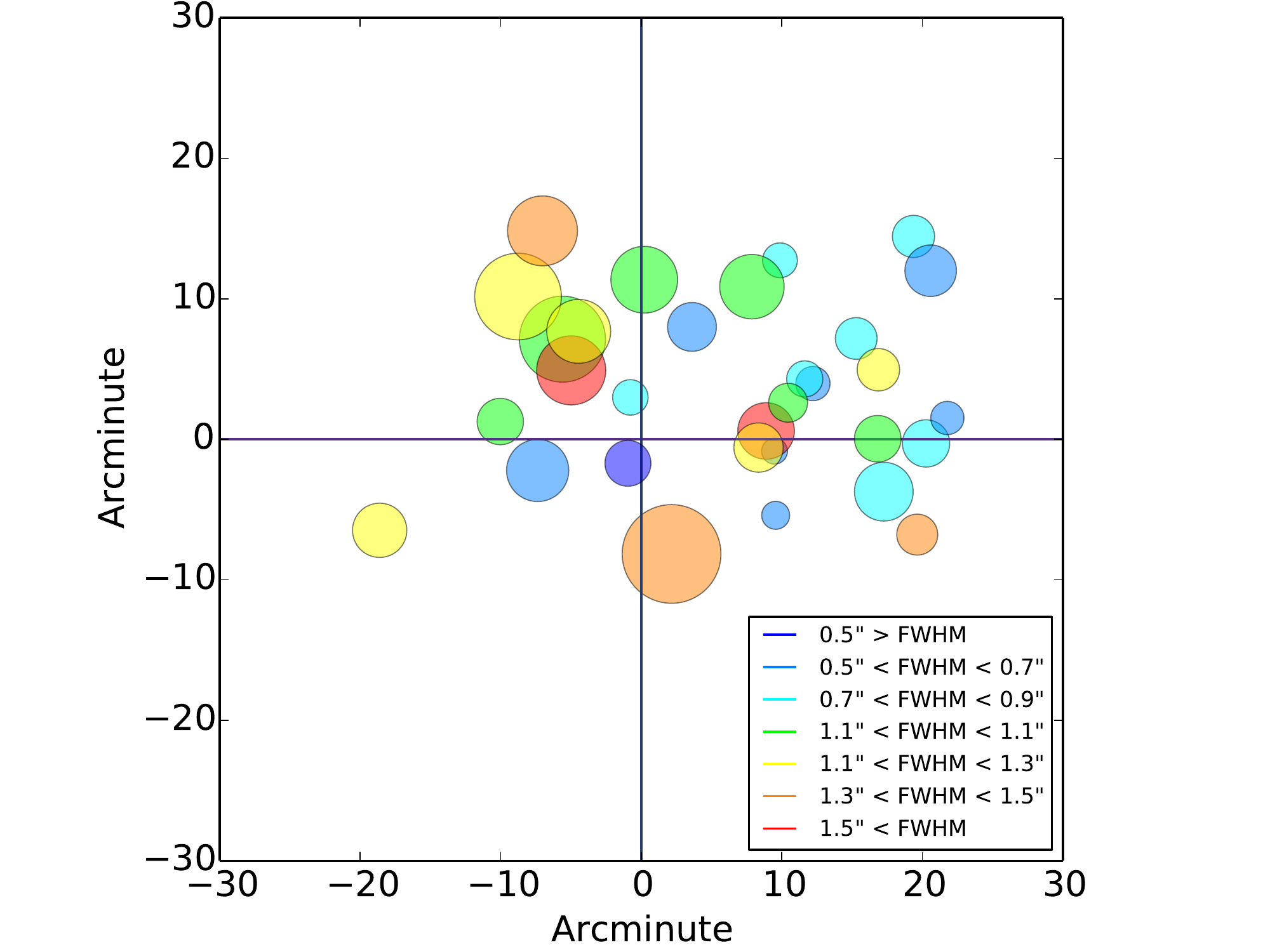}
\centering
\caption{Effect of the  location of the reference star used for the r--band PSF modeling  on the ghost strength. 
All stars used for the PSF modeling are plotted together.
 The disk position corresponds to the spatial offset of each star  with respect to the center  of  the stacked images. 
Its size is proportional to the  intensity of the  star halo determined at a radius of  100\arcsec,  after a normalization of the profiles.
 Its color indicates the FWHM of the star (with a color code given in the legend),  to take into  account  the  possible  additional effect of the
seeing.  }
\label{fig:StarPositions}
\end{figure}

 We choose here an hybrid method. To create the PSF model, we build
 the inner profile by stacking many faint stars on the field and the
 outer profile by manually modeling the ghosts of a bright neighbor
 star; we then merge the inner and outer profiles.  The inner parts of
 the PSF are generated using SExtractor \citep{bertin96} and PSFex
 \citep{bertin11}.  Since the ellipticity, and full width at half
 maximum (FWHM) of the PSF vary across the field of view, the
 reference stars were chosen within a box of 20\arcmin x20\arcmin\, centered
 on the target galaxy. This generally allows us to build the inner PSF model up to a radius of
$\sim$6.5\arcsec. The radius is reduced when the number of  stars available for the PSF model with PSFex 
is insufficient and the signal to noise too low.

 For the outer regions of the PSF, we directly exploit the depth of
 the MegaCam images that allows us to model the light profile of
 bright (saturated) stars up to several arcmin. 
 In most cases, the closest bright
 star to the target galaxy  exhibiting prominent  ghosts was
 used for the PSF fitting procedure.  The positions of the various selected stars are shown in Figure~\ref{fig:StarPositions}.
  The wings of the PSF are
 computed with a similar  manual method as the one used to remove the stellar ghosts (see
 Section \ref{SubofStel}), i.e. using a series of annuli with discrete
 intensity. 
 One difference is that the fitting procedure was continued to a smaller radius, 
 i.e.  $\sim$4.5\arcsec.
 
 With this method the outer PSF is azimuthally averaged.  Getting away from the
 center of the MegaCam field, the ghosts are less and less   centered on the
 star. For this reason,  in our star profile fitting procedure, the center of the  annuli are allowed to be shifted.
To build our PSF model, the annuli were however artificially re-centered.
Indeed, the vast majority of our target  galaxies are located close to the image center at a position where the ghost halo 
is centered on the star.
With that approximation, we make sure that  the deconvolution of the galaxy 
is carried out with a PSF with a shape  close to the expected one.

\begin{figure*}[t]
\includegraphics[width=0.8\textwidth]{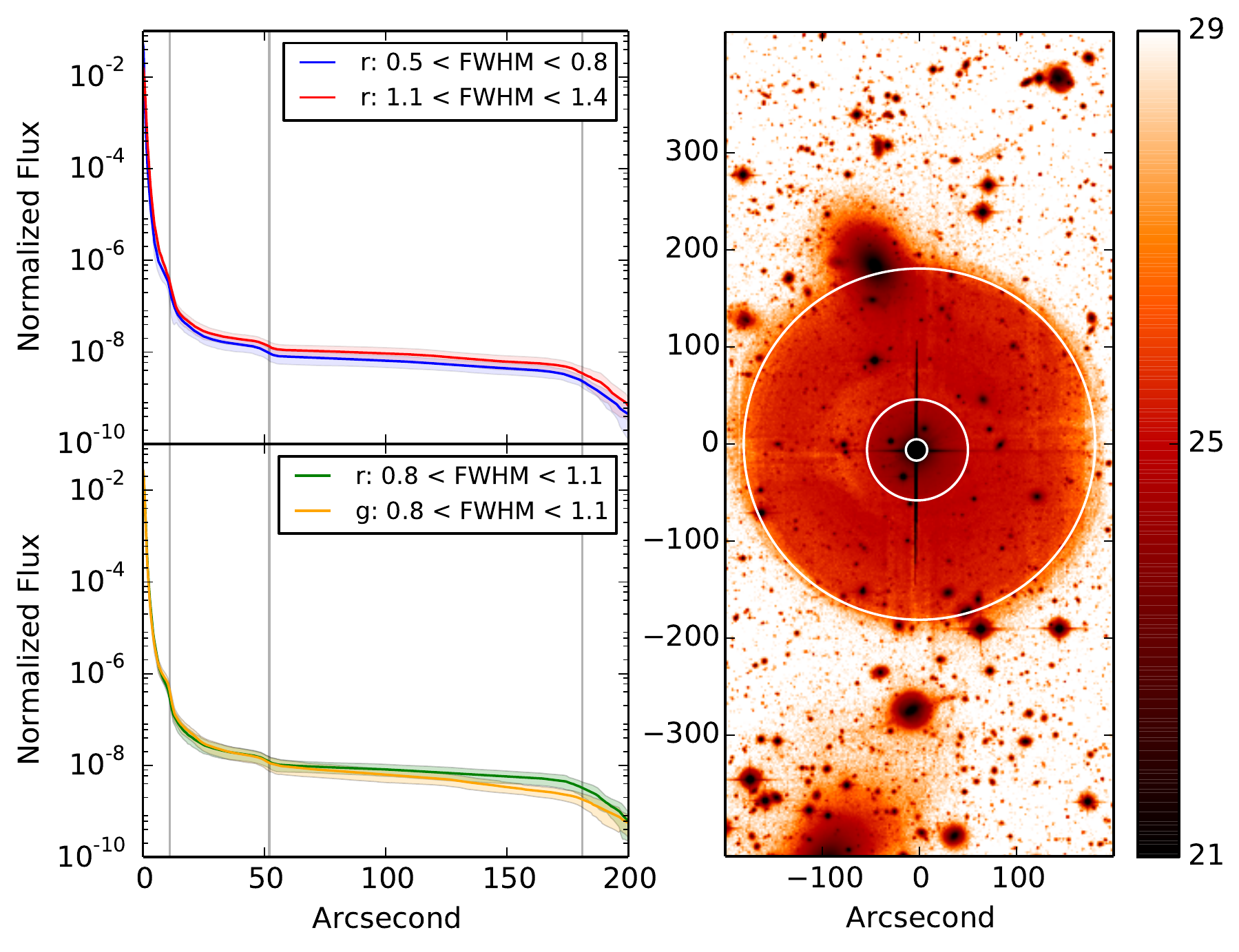}
\centering
\caption{Mean normalized profile of the  r-band PSF model profiles for two bins of seeing 
  \textit{(top-left)};  mean profile of the g- and r-band profiles for a  specific seeing bin  \textit{(bottom-left)} and the r-band surface
  brightness map of a typical stellar ghost \textit{(right)}. The standard deviation of each profile is shaded. The
  surface brightness scale in mag.arcsec\textsuperscript{-2} is given
  to the right. The map is blurred with a Gaussian (sigma =
  0.55\arcsec) to enhance the faint regions. The vertical black lines
  in the plots to the left indicate the radii with sharp breaks in the
  profiles. Their positions on the surface brightness map are
  indicated by the white circles.  }
\label{fig:PSFProfiles}
\end{figure*}

Merging the inner and outer profiles of the PSF model requires a
normalisation in the overlapping region. The normalisation factor is
computed as the median ratio of the pixel values in a common region
with an angular size of $1.1\arcsec$ (i.e. 6 pixels).

Figure~\ref{fig:PSFProfiles} presents the typical  profiles of the PSF models, for various seeing conditions 
and the two g and r bands. 
All PSF models were normalized by their estimated  flux at 3\arcsec, and  combined within each seeing bin, and a standard
deviation was computed.
 The standard deviation curves shown in the figure indicate that the profiles are not uniform for a given seeing, band and star magnitude. A typical error of 30\% at  radius of 100 \arcsec  is measured.
 This could first be  due to uncertainties in the merging process of the inner and outer profiles of the PSFs, and to the spatial variations of the PSF wings. The impact of the latter effect is illustrated in Figure~\ref{fig:StarPositions} and further discussed in Section~\ref{LimitonPSF}.

 Though the  intensity scaling used in our figures makes the ghost halos appear prominent,  their contribution to the total flux is very limited: the  flux integrated between $\sim$50-200\arcsec\ accounts for only 1.6\%; 98\% of the flux is concentrated within the inner 10\arcsec.

We found that  for a given seeing range, at radius above $75\arcsec$,  the r--band reflected  light
becomes more intense than in the g--band. This is mainly due to the
CCDs being more reflective in the r-band. This wavelength dependence
of the PSF creates artificial reddening of the outer profiles of
galaxies.

More surprisingly, for a given band,  the PSF models corresponding to worse seeing
conditions have more prominent halos. The difference is significant: about 45\% at 100\arcsec, while the standard deviation per seeing bin is $\sim$ 30\%. We checked that a normalization of the flux with an aperture of 10\arcsec, instead of 3\arcsec\ does not change the results.

Understanding the origin of these trends is beyond the scope of this study and will be discussed in another paper presenting a physical rather than empirical modeling of the PSF. 
In any case, the strong wavelength, seeing and position dependence of the  PSF wing intensity  presented here  further illustrates the need of having specific PSF models for each observation
condition.  However, as described in Section \ref{SubofStel}, obtaining a full  PSF model with our manual empirical method is  time consuming.  For that reason,
instead of creating one PSF for each image of our large sample, we
have generated one PSF for each band and for different seeing conditions, ranging from 0.5\arcsec\, to 1.7\arcsec.\footnote{Note however 
that, for the vast majority of the MATLAS fields, the r--band seeing ranges between $0.5\arcsec$  and $1\arcsec$. }
A database of 71 PSF models was then generated.

\subsection{Image deconvolution} \label{Deconv}

To remove the ghost halos due to scattered light, we have directly
deconvolved our images using the very large kernels (PSFs) presented
in Section~\ref{ObtainPSF}. The deconvolution technique on large
images with large kernels have two downsides: (1) it requires a large
computation time even for low numbers of iterations (2) the
deconvolved images exhibit irregularities, i.e.  wiggles and noise on
small scales. This second issue is just intrinsic to the technique and
unavoidable as the PSF model can never be perfect. We have however
dramatically reduced the computer time using specific python operators
and solvers for high-performance computing.

The deconvolution method described here uses open source PyOperators,
in particular convolution operators and equation solvers that were
optimised to handle large matrix calculations. This allows the process
to reach high iterations in relatively short time.  More specifically,
the code uses a conjugate gradient method; the maximum iterations and
the tolerance level for the residual are controlled by the user.

Note that the cost of solving the minimization equations directly by a Cholesky decomposition
 would be prohibitive. The iterative preconditioned conjugate gradient method is 
 more efficient.

The noise in the deconvolved images due to the noise in the original
images plus the imperfections in the representation of the PSF by the
kernel is regulated by a regularisation parameter. In addition, a
normalisation parameter\footnote{Determined for each image as 1
  divided by the square of the sigma of the sky background.} is given
to help reaching the solution in low number of iterations.

\subsection{Determination of the galaxy profiles} \label{LProfiles}

To derive the galaxy profiles, we have used the following routine.
First, the small objects (faint stars and background galaxies) were
subtracted from the images, after having been identified by SExtractor
and replaced by values of the surrounding background.  Bad regions
(including prominent CCD gaps, galactic cirrus, bright objects etc.)
were then manually masked.  Initial photometric parameters were
estimated by GALFIT, and the modeling of the galaxy itself was made by
the \textit{ellipse} task in \textit{IRAF}.  In the ellipse fitting
procedure, the position angle and ellipticity parameters were set
free, except for galaxies with prominent bar-like structures, dust
lanes and bright clumps.  The ellipse procedure usually breaks at
locations were bright clumps within the galaxy or fine structures
(tails, shells) become prominent.  To force the determination of the
profile farther out, a maximum semi-major axis was given.  The
profiles were however truncated at a fixed level of the measurement
error, as described in Section~\ref{ResonETG}.

\begin{figure}[t]
\includegraphics[width=8.8cm]{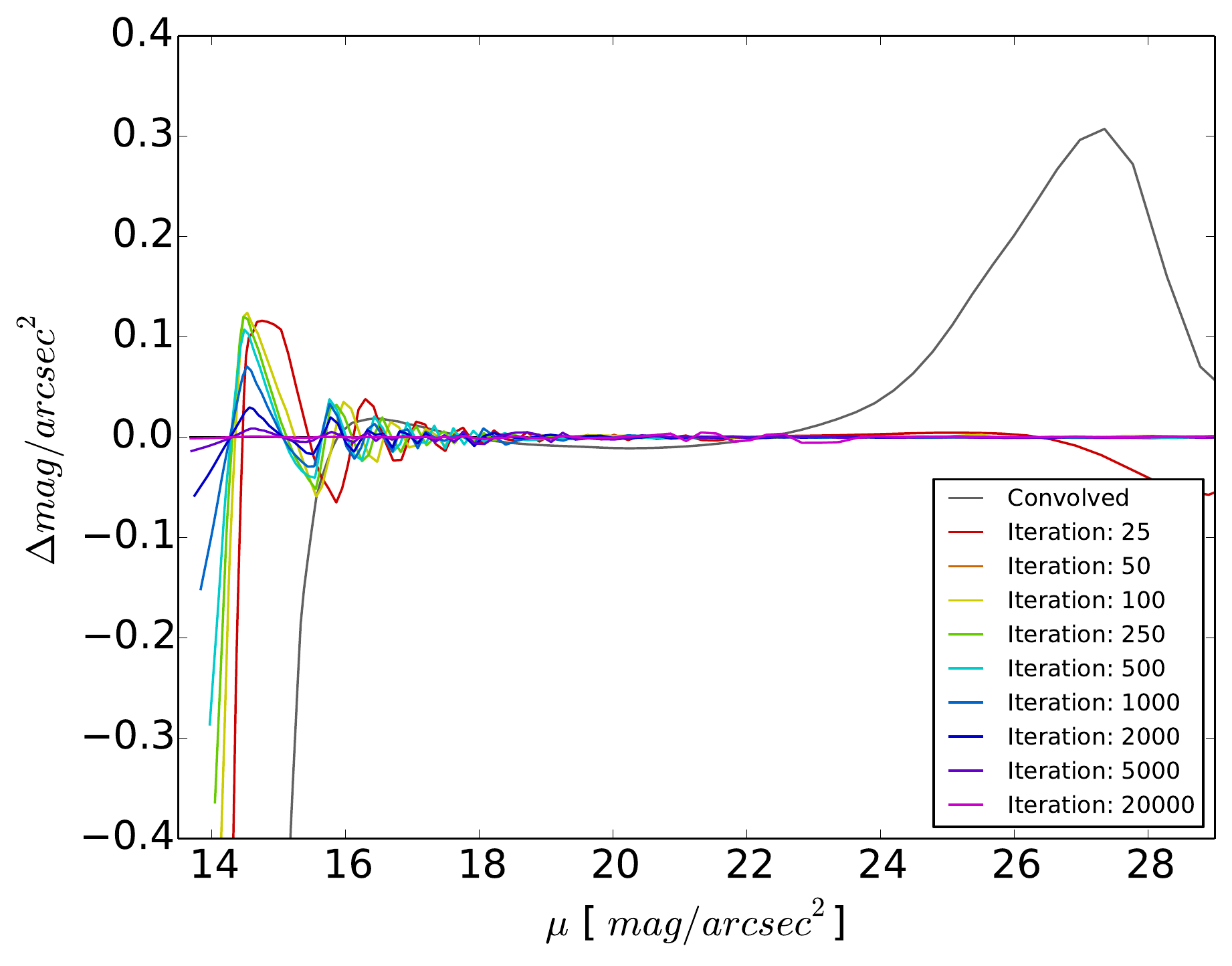}
\centering
\caption{The effect of the PSF on the  profiles of simulated
  galaxies. The black curve shows the surface brightness difference
  between the convolved and original profiles of the simulated galaxy
  as a function of local surface brightness. The image is convolved
  with a r--band MegaCam PSF model and a FWHM equal to 0.81\arcsec. The
  offset becomes significant above 25 mag.arcsec\textsuperscript{-2}
  and reaches 0.3 mag.arcsec\textsuperscript{-2} at the 27
  mag.arcsec\textsuperscript{-2} isophote. The colored curves show the
  difference between the deconvolved and original profiles for
  different number of iterations (given in the box).}
\label{fig:SimIteration}
\end{figure}

The large scale sky background variations are the principle sources of
systematic uncertainties in the light and color profile
determination. Scattered light due to Galactic cirrus or clustered
stars generate large fluctuations across the field. The determination
of the average background around the target galaxy and its error is
thus not straightforward. The code we have developed for this process
samples the background in 190 homogeneously distributed circular
regions (150 pixel diameter) around the galaxy. The sky value and
standard deviation is derived for each of these regions by IRAF's
\textit{centroid} sky fitting algorithm. Regions polluted by CCD gaps,
stellar ghosts or strong cirrus contamination are identified based on
the value of the standard deviation and rejected in a two iteration
process. For the remaining regions, the robust mean and sigma for the
sky is calculated with the \textit{biweight\textunderscore mean} in
IDL \citep{idlastro}.

The local errors are directly given by the \textit{ellipse} fitting
task.  The error bars attached to the light and color profiles combine
quadratically the systematic and local errors\footnote{Note that in
  the literature, the error on the sky background is usually
  neglected, resulting in much smaller error bars than those shown in
  this paper.}.

\begin{figure}[t]
\includegraphics[width=8.8cm]{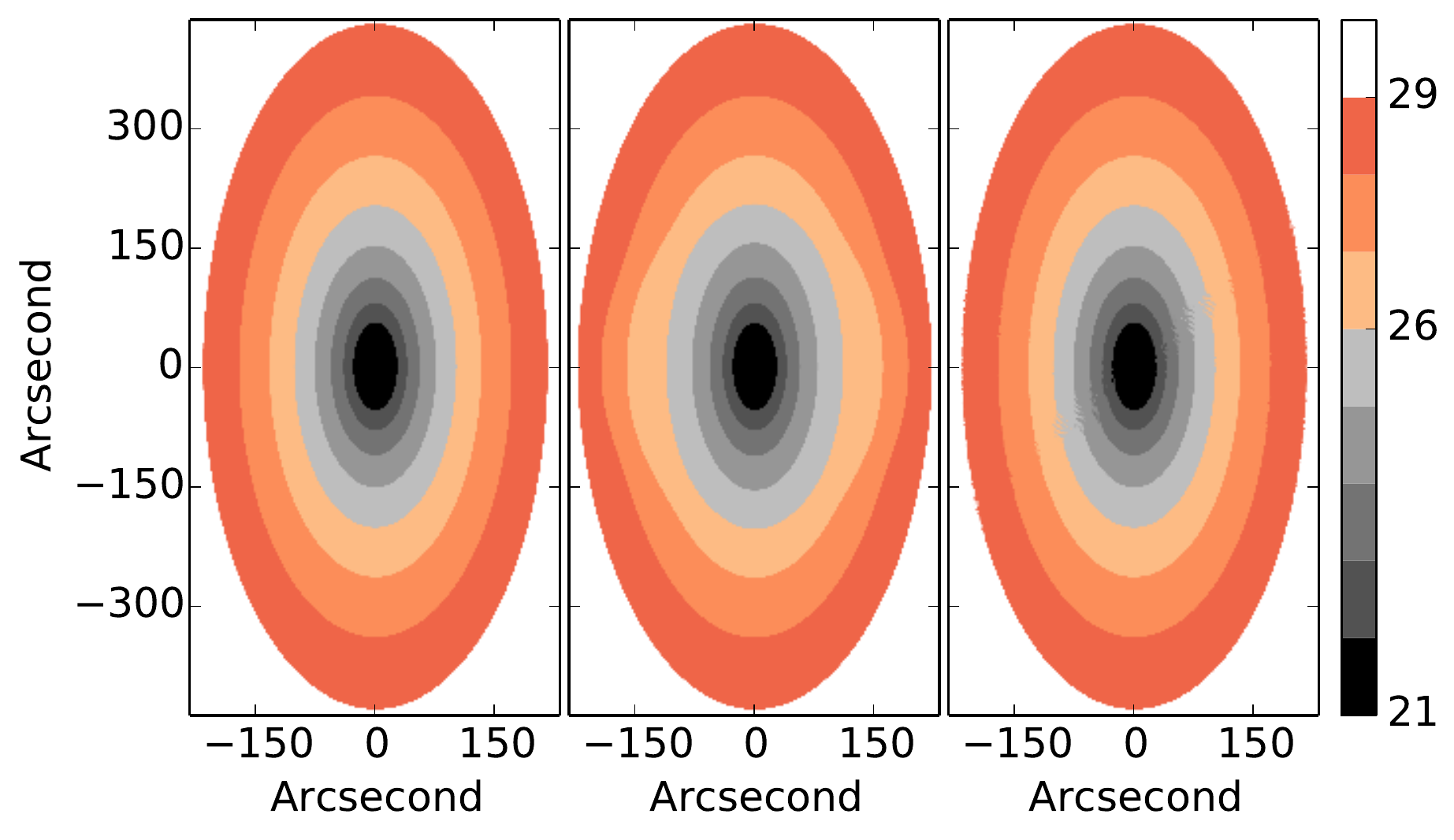}
\centering
\caption{Surface brightness maps of the original \textit{(left)},
  convolved \textit{(middle)} and deconvolved \textit{(right)} images
  for the simulated galaxy. The color bar to the right represents the
  surface brightness level. The image is convolved with a PSF in
  r-band with FWHM=0.81\arcsec', and the deconvolved image is obtained
  after 20000 iteration steps by the solver.}
\label{fig:SimVisual}
\end{figure}

\section{Results} \label{Results}

In this section we present the deconvolution results on simulated
images of galaxies and real images of NGC\,3489, NGC\,3599 and
NGC\,4274. The analysis of the full sample will be presented in a
future paper.

\subsection{Simulations} \label{ResonSim}

The deconvolution technique was tested on convolved images of
simulated galaxies. A single component galaxy model was created with
GALFIT \citep{peng02,peng10} and inserted on the images, leaving
enough padding to avoid any boundary effects in the convolution and
deconvolution processes.  The simulated galaxy has the following
characteristics, typical of those of the early-type galaxies in our
sample (see Table~\ref{tab:PropGal}).

\begin{itemize}
  \item S\'ersic Index: 4
  \item Integrated Magnitude: 12.1  within a 5\arcsec\,aperture
  \item Effective Radius: 22.3\arcsec
  \item Ellipticity: 0.5
\end{itemize}

The intrinsic simulated galaxy image was convolved by the convolution
tool in PyOperators with a PSF obtained from a real r-band image. Its
FWHM of 0.81\arcsec\, is close to the average seeing in r--band for the
whole MATLAS survey. The deconvolution was performed on the convolved
image with the code described in Section~\ref{Deconv}. No
regularisation parameter was set as the simulated image is
noiseless. The deconvolution was tested with different number of
iterations, ranging between 25 to 20000.
Figure~\ref{fig:SimIteration} shows that convergence in the outskirts
of the galaxy is already reached with a small number of iterations
($\sim$100). As expected, recovering the true inner profile requires
however a much higher number of iterations.

Figure~\ref{fig:SimVisual} presents the projected images of the
intrinsic, convolved and deconvolved galaxy. It illustrates how light
scattering not only change fluxes but also affects the outer isophotes
of the galaxy: the halo gets artificially roundish, especially between the 26th and 27th  mag.arcsec\textsuperscript{-2} isophotes. This effect can
vary from galaxy to galaxy depending on the size, ellipticity and
brightness of the nucleus.  As shown on the figure, the deconvolution
technique is able to recover the intrinsic shape of the modelled
galaxy.

\subsection{Observations: NGC\,3489, NGC\,3599 and NGC\,4274} \label{ResonETG}

\begin{figure*}
\includegraphics[width=18cm]{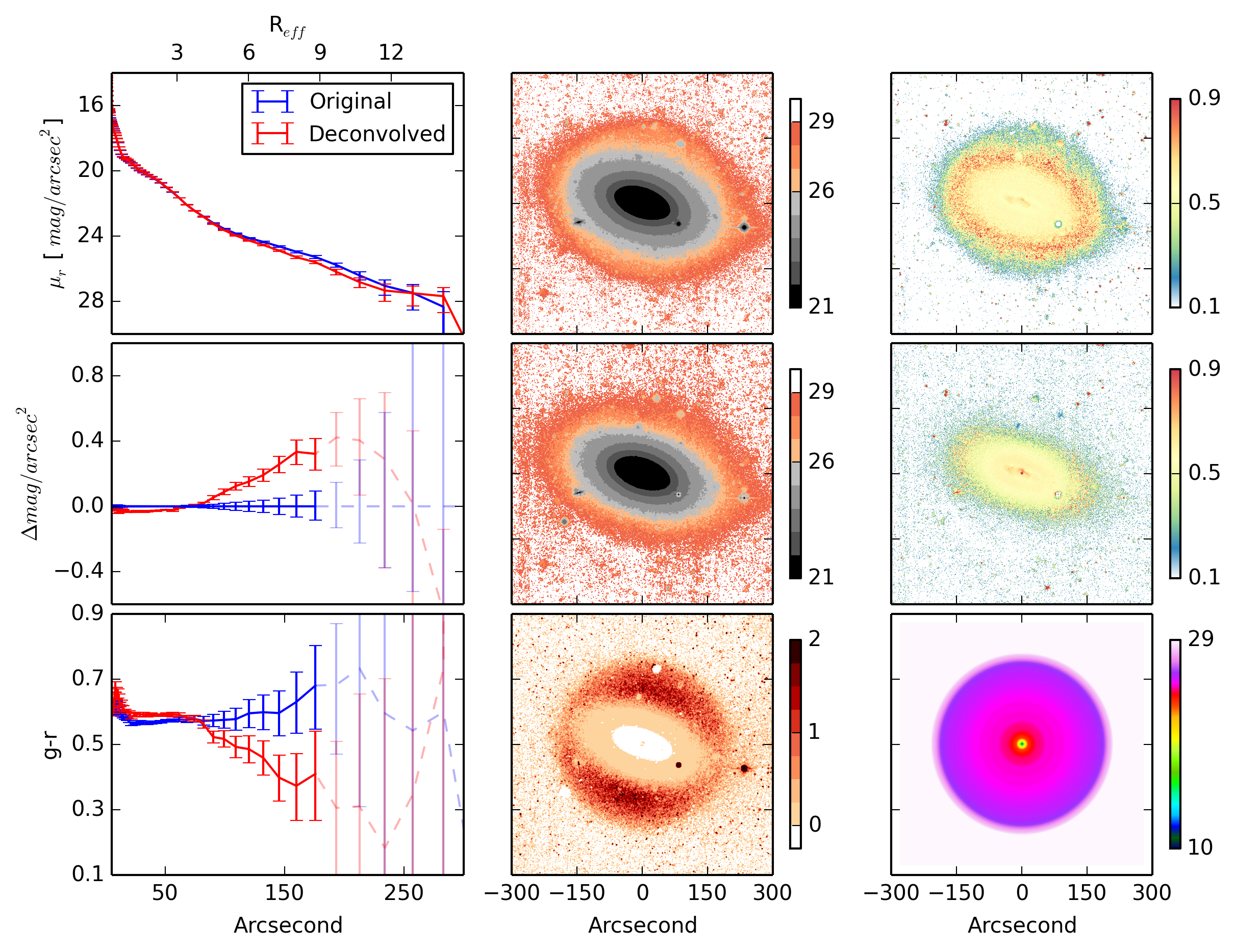}
\centering
\caption{The deconvolution technique applied to NGC\,3489. \textit{Left Column:} Original and deconvolved galaxy profiles in r-band \textit{(top)}, the difference between the profiles \textit{(middle)} and original and deconvolved g-r color profiles \textit{(bottom)} as a function of semi-major axis. \textit{Middle Column:} Original \textit{(top)} and deconvolved \textit{(middle)} surface brightness maps in r-band, and the difference between the maps \textit{(bottom)}. \textit{Right Column:} Original \textit{(top)} and deconvolved \textit{(middle)} g-r color maps and the r--band surface brightness map of the PSF \textit{(bottom)} that was used for the deconvolution. 
  All maps except PSF were smoothed with a Gaussian
  kernel of sigma = 0.55\arcsec\, to enhance the visibility of the faint
  regions.}
\label{fig:NGC3489}
\end{figure*}

\begin{figure*}
\includegraphics[width=18cm]{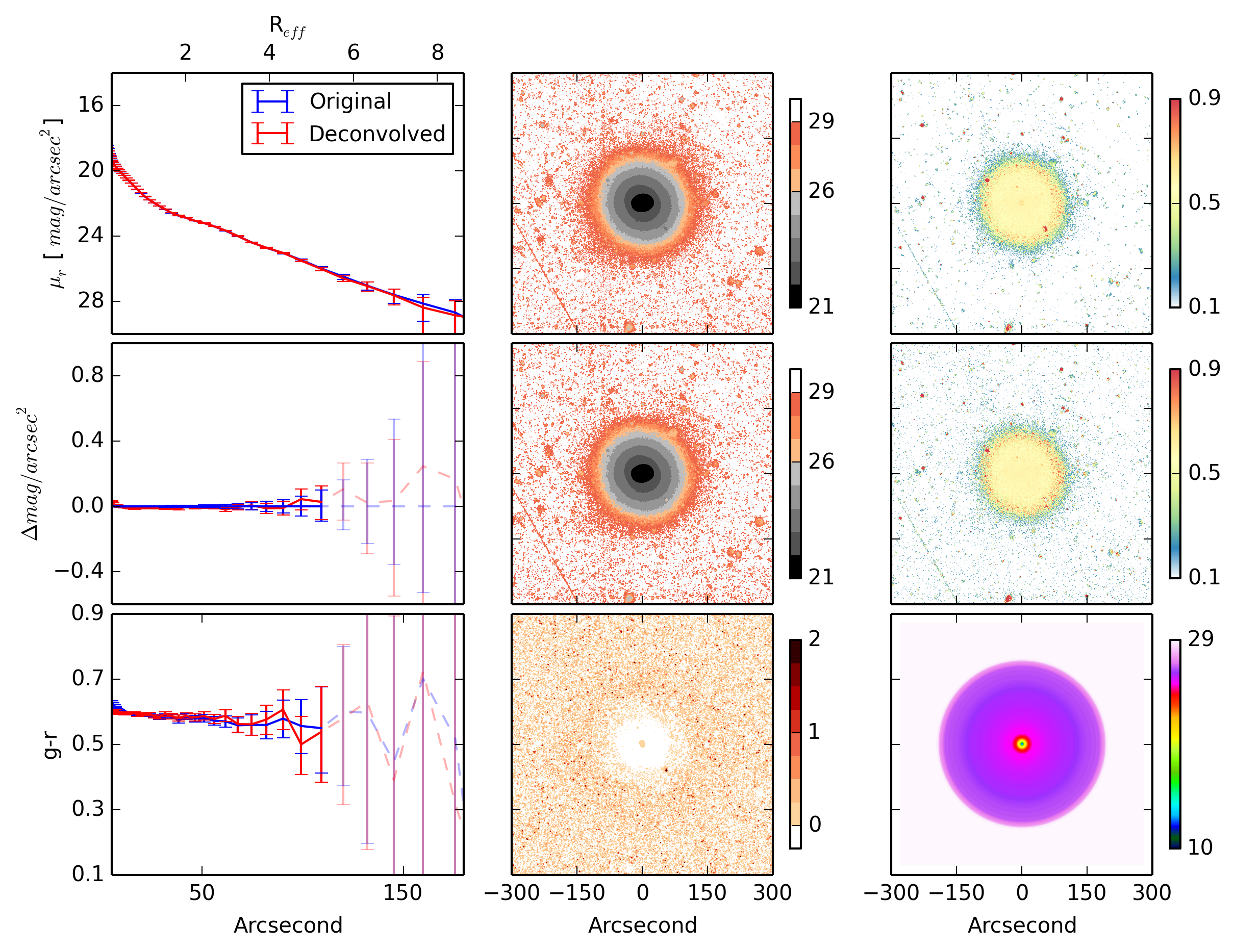}
\centering
\caption{The deconvolution technique applied to NGC\,3599. See caption
  of Figure \ref{fig:NGC3489} for details.}
\label{fig:NGC3599}
\end{figure*}

\begin{figure*}
\includegraphics[width=18cm]{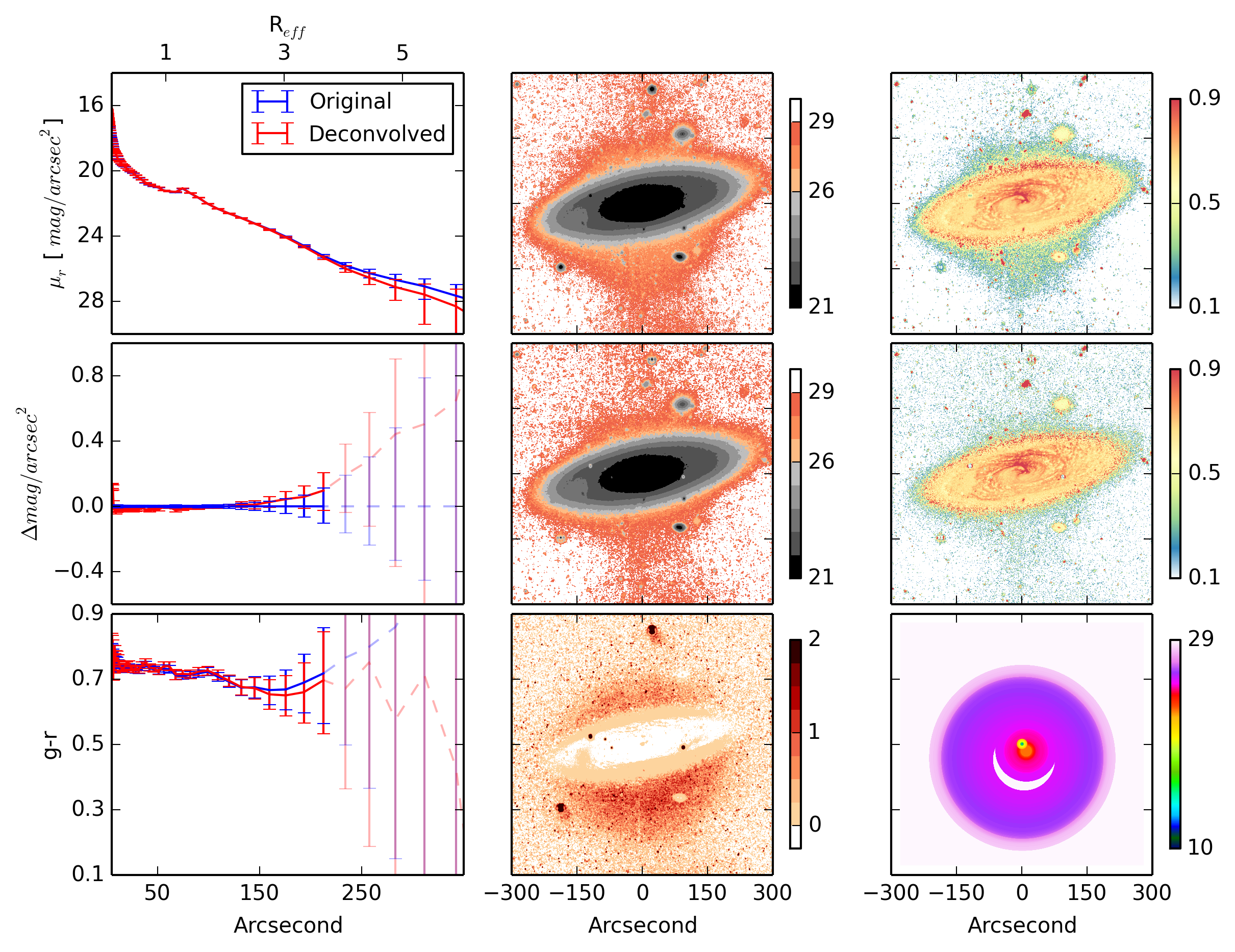}
\centering
\caption{The deconvolution technique applied to NGC\,4274. See caption
  of Figure \ref{fig:NGC3489} for details.}
\label{fig:NGC4274}
\end{figure*}

The deconvolution technique was then applied to the images of real
galaxies NGC\,3489, NGC\,3599 and NGC\,4274, chosen for being
representative of the variety of objects in the MATLAS survey.  Their
properties are listed in Table~\ref{tab:PropGal}.

\begin{table}[h]    \begin{threeparttable}
	\centering
		\caption{The properties of the tested galaxies.}
	\label{tab:PropGal}
	\begin{tabular}{@{}lccc@{}}
	\toprule
                  & $R_{Eff}$ & Central Brightness & Integrated Magnitude \\ \midrule
	NGC\,3489 & 19.9 & 14.3 & 11.7 \\
	NGC\,3599 & 20.8 & 16.2 & 13.5 \\ 
	NGC\,4274 & 58.0 & 16.6 & 13.1 \\
	\bottomrule
	\end{tabular}
	\begin{tablenotes}
	      \small
            \item Notes: Column (2): Effective radius taken from the
              literature in arcseconds
              \citep{cappellari2013a,vika2013}; Column (3): Central
              brightness of the light profile in r-band in mag.arcsec\textsuperscript{-2}; Column(4):
              Integrated magnitude within 5\arcsec\, (aperture) in
              r-band.
 	   \end{tablenotes}
	  \end{threeparttable}
\end{table}

NGC\,3489 exhibits a prominent ghost that can be seen directly in the
surface brightness and color maps as a roundish, reddish, structure
with a radius of 3.5 arcmin, typical of that surrounding bright stars
(see Figure~\ref{fig:NGC3489}).

We performed the deconvolution on the g- and r-band images of
NGC\,3489, introducing this time a regularisation parameter. The technique successfully
eliminates the artificial red halo after 500 iterations \footnote{For
  real images, a higher number of iterations does not help to get rid
  of the wiggles generated by the deconvolution technique in the inner
  regions. Noise in the initial images and imperfections in the PSF
  modeling combine to prevent any efficient correction. For our
  purpose - the correction of the outer regions from light scattering
  -, these artificial irregularities in the deconvolved image are not
  a problem.}.
 
The profile in the r-band is corrected by up to $\sim$0.4
mag.arcsec\textsuperscript{-2} at the 26
mag.arcsec\textsuperscript{-2} isophote, corresponding to about 9
$R_{Eff}$. Because of the wavelength dependence of the PSF, the color
profile is also affected by the light scattering.  At the 26th
mag.arcsec\textsuperscript{-2} isophote, the deconvolution makes the
g-r color bluer by 0.2 mag. This results in an overall change of the
color gradient: while the initial profile indicated an (artificial)
reddening, the corrected one shows instead a bluing.
 
The change in the color profile becomes significant, i.e. larger than
the error bar, beyond 5 $R_{Eff}$.
The error bars of the color profiles are calculated by taking the square root of the sum of the square of each band's error. \\

NGC\,3599 is representative of the lowest mass galaxies in  the
MATLAS/ATLAS$^{\sc 3D}$ sample  (see Table~\ref{tab:PropGal}). 
Results for this galaxy are shown in  Figure~\ref{fig:NGC3599}. Both 
the light and the color profile show almost no change after deconvolution 
and the possible correction in the outermost part stays within the error 
bars due to the low signal to noise level in this region.\\

Finally the technique was tested on a late-type galaxy. LTGs that are
not seen face-on have higher ellipticity and thus a shape that differs
from the roundish one characteristics of the ghost halos. The
scattered light is more likely to become dominant along the minor
axis. LTGs are thus good cases to test the deconvolution technique.
Although the MATLAS survey targets lenticular and elliptical galaxies,
a number of spiral galaxies are present in the one square degree
MegaCam field centred on the ETGs. At their location, often away from
the center of the field, the PSF wings may be very different from the
inner ones, and in particular the ghost halos may no longer be
concentric. This is an opportunity to test the deconvolution method
for this more complex configuration. We present here as a test case
the LTG NGC\,4274, located in the field of the ETG NGC\,4278.  An
accurate PSF modeling was performed, and contrary to the previous
cases, the centres of the halos were allowed to change positions.  The
results are shown in Figure~\ref{fig:NGC4274}. The deconvolution
removed most of the original prominent ghost around the galaxy (but
due to the position of the galaxy far from the center, not centred on
it), but a faint residual is still visible, likely the result of the
imperfect modeling of the PSF.


\section{Discussion} \label{Discussion}

In this section, we discuss the limitations of the deconvolution
method, in particular due to the imperfect PSF modeling and
saturation.  The overall importance of scattered light in the MATLAS
survey is briefly addressed.

\subsection{Limitations of the method} \label{LimitonPSF}

The ability to recover the original profile in simulated images, and
the removal of the ghosts around galaxies in real images testify the
reliability of the deconvolution technique to achieve our primary
goals: studying the light and color profiles of galaxies at large
radius.  However, as stressed before, in all convolution/deconvolution
techniques, having a perfect knowledge of the PSF is critical. Our
modeling of the PSF based on stars located within the science frames
has some limitations.

First, the outer profile of each PSF is derived from a single bright
star in the image that is not located at the exact position of the
galaxy; it is in most cases selected as the closest bright star having a prominent
stellar ghost. The PSFs are subsequently built for the center of the
images and the reflections are modeled as concentric halos. The latter
hypothesis is justified at the position of the target galaxy, close to
the center of the MegaCam field, where no important decentering is
observed. As illustrated in Section~\ref{ResonETG} with the case of
the off-centered galaxy NGC\,4274, even when allowing a decentering
for the PSF model, the ghost subtraction is not perfect.

Besides, the PSF has been derived from stacked images and not from the
individual ones. Our observing strategy requires to implement large
offsets between them, up to 14 arcmin.  So, as illustrated in Figure~\ref{fig:GhostPositions},  the shape of the PSF, and
associated ghost halos,  differ from one image to the
other. Thus in principle, a PSF model should be built for each image,
and the deconvolution made before stacking using this specific
PSF. This procedure would however highly increase the interaction and
computation time making our largely manually method unpractical for
the whole sample.  As a tradeoff, a proper parametrisation of the
ghost properties, and determination of the parameter spatial
variations using linear regressions could be done \citep{slater09}. An
even better method is to derive the PSFs with a physical model of the
principle internal reflections in the camera. It would allow us to
predict the PSF shape at each pixel. There are on-going efforts to
achieve this for the MegaCam camera.

One should also note that, in practice, the individual images most
often lack signal to build the outer wings of the PSFs with the method
described in Section~\ref{ObtainPSF}, and given the low S/N in the
galaxy outskirts, the deconvolved image would be extremely noisy.  The
stacking procedure we used blurs the differences due to the large
offsets. However it does it the same way for the stars used to derive
the PSF and the target.

\subsection{The importance of scattered light in the MATLAS survey} \label{Estimation}

The results of both simulations and observations show that the
scattering should significantly vary from galaxy to galaxy. In the
MATLAS sample, we expect to see a significant effect for some of the
galaxies and almost none for others, depending on the brightness of
the nucleus and also on galaxy size.  When the stellar light and
signal-to-noise decrease, the correction for the scattered light
becomes smaller than the error bars.

\begin{figure}[t]
\includegraphics[width=8.8cm]{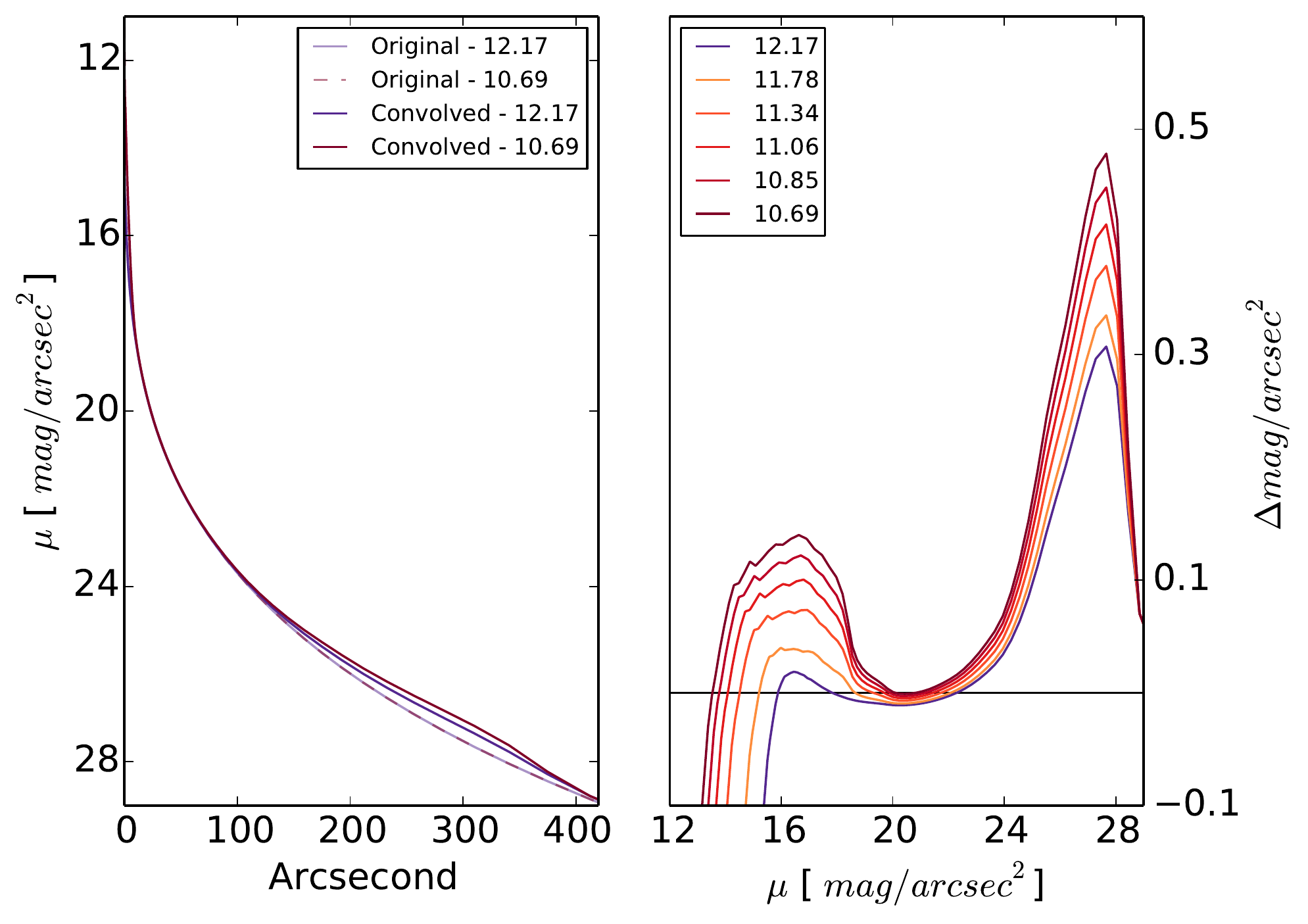}
\centering
\caption{Surface brightness profiles of the PSF-convolved and the original simulated galaxies \textit{(left)} and difference between them as a function of surface brightness \textit{(right)} for different central magnitude. Each line represents one of the
  simulated galaxies. The integrated magnitude within 5\arcsec\,
  (aperture) is written in the legend for each galaxy.}
\label{fig:SimGalaxies}
\end{figure}

We have created a simulated galaxy to test how the brightness of
nucleus determines the amount of scattered light. The galaxies'
properties are selected as described in Section~\ref{ResonSim}. New
galaxy models with higher central surface brightness are subsequently
produced keeping the same light profile beyond r=5\arcsec\, and
steepening the inner profiles.  All these images are then convolved
with a real r-band PSF. The results are shown in
Figure~\ref{fig:SimGalaxies}. Brighter cores indeed create more
artificial flux which, for the model tested here, starts to be
significant, i.e. above 0.1 mag.arcsec\textsuperscript{-2}, beyond the
~24 mag.arcsec\textsuperscript{-2} isophote. The effect peaks at the
27 mag.arcsec\textsuperscript{-2} isophote, and decreases at lower
surface brightness, i.e. at large radius. This is due to the fact that
for our model, the galaxy is more extended than the primary ghost (of
3.5 arcmin radius).

Conversely, if for whatever reason, including the
presence of dust lanes, the nuclear light is dimmed, the ghosts become
less prominent.  Saturation is a special case for which the flux on
the image is artificially dimmed while the light of the nucleus
arriving on the detector and level of internal reflections is
unchanged. Here, the image should have a ghost contamination higher
than anticipated from the measured central brightness. In other words,
since the deconvolution process corrects the scattered light effect in
the outer region based on the bright core, deconvolution of saturated
sources leads to underestimated corrections.  About 75\% of the
galaxies in the MATLAS long exposures are saturated and short
exposures have been acquired for all of them to recover the true inner
profiles of the galaxies. The observations and the method of merging
long and short exposure images to prevent saturation will be described
in a future paper, together with the deconvolved light and color
profiles of all MATLAS galaxies.

Note, however, that saturation will not necessarily significantly
affect the reconstructed light profile for all galaxies.  The effect
is expected to be worse for relatively small low luminosity galaxies
with a bright saturated nucleus. Such objects are actually rare in the
ATLAS$^{\sc 3D}$ sample.

\section{Conclusions} \label{Conclusions}

In this paper we have addressed the instrumental scattering effect in
deep CFHT/MegaCam images. Our results are however relevant for the
many cameras suffering from instrumental reflections (such as the one
equipping the Sloan or ESO VST).  These reflections produce wings --
ghost halos -- in the PSF, the importance of which depend on the
wavelength and position in the field of view.  With MegaCam, they are
worse in the r--band. The wings of the PSF generate themselves ghost
halos around the galaxies, that may be directly visible on the images
as round, reddish, disks. The scattered light affects the shape of the
light and color profiles of galaxies. Our simulations show that the
effect is worse between the 24th and 28th
mag.arcsec\textsuperscript{-2} isophote, a region of prime importance
as it accumulates tidal debris and thus trace the last accretion
events of galaxies.

We have presented a method, based on a direct deconvolution of the
image, which is able to efficiently remove the ghost halos in a
relatively low number of iterations.  It requires a proper modeling of
the PSF, made combining the inner profile of stacked faint stars in
the field, and the outer profile of a bright (saturated) star close to
the galaxy. We have built a PSF database covering different seeing
conditions, bands and locations on the MegaCam field of view.  With
respect the model-convolution technique also used to remove artificial
halos in deep images, the deconvolution technique has the advantage of
not requiring a proper, multi-component, modeling of the galaxy. It is
thus much faster for large samples of complex galaxies, such as those
studied as part of the MATLAS deep imaging survey.  Intensive tests of
the method have been made on simulations and on real galaxies
presenting a variety of morphologies and sizes. We have presented here
the results for three test cases: NGC\,3489, NGC\,3599 and NGC\,4274.

The saturation in the cores of some of the galaxies in principle
averts the proper removal of the ghosts.  A follow-up short exposure
program has been completed to allow us to carry out a proper
analysis. The full analysis of the sample will be presented in a
following paper.

\begin{acknowledgements}
We thank the referee for his carefully reading of the paper and for having pointing out
 some issues that  required some further analysis. 
  The paper is based on observations obtained with MegaPrime/MegaCam,
  a joint project of CFHT and CEA/DAPNIA, at the Canada-France-Hawaii
  Telescope (CFHT), which is operated by the National Research Council
  (NRC) of Canada, the Institute National des Sciences de l'Univers of
  the Centre National de la Recherche Scientifique of France, and the
  University of Hawaii.  All observations were made as part of the
  service mode offered by the CFHT. We are grateful to the queue team
  for their dedication.  We warmly thank C. Mihos, N. Regnault,
  C. Sandin, I. Trujillo and L. Kelvin for the multiple discussions on
  ghost and scattered light effects.  We are grateful to the CFHT
  staff for they dedication to the data acquisition.

\end{acknowledgements}

%
%

\bibliographystyle{aa}
\bibliography{PaperDeconv}

\end{document}